\documentclass[11pt,a4paper]{article}

\usepackage[margin=1in]{geometry}
\usepackage{tabularx}
\usepackage[T1]{fontenc}
\usepackage{lmodern}
\usepackage[utf8]{inputenc}
\usepackage[protrusion=true,expansion=false]{microtype}
\usepackage{amsmath,amssymb}
\usepackage{booktabs}
\usepackage{array}
\usepackage{longtable}
\usepackage{tabularx}
\usepackage{graphicx}
\usepackage{xcolor}
\usepackage{enumitem}
\usepackage{listings}
\usepackage{caption}
\usepackage{float}
\usepackage{needspace}
\usepackage[numbers,sort&compress]{natbib}
\usepackage{hyperref}

\newcolumntype{Y}{>{\raggedright\arraybackslash}X}

\hypersetup{
  pdftitle={HEPLocalAgent: Running Collider Simulations from Plain-Language Requests on Your Own Computer},
  pdfauthor={Aadarsh Singh and Sudhir K. Vempati},
  colorlinks=true,
  linkcolor=blue!60!black,
  citecolor=blue!60!black,
  urlcolor=blue!60!black
}

\newcommand{\agentname}{HEPLocalAgent}

\newcommand{\proc}[1]{\texttt{#1}}
\newcommand{\agentversion}{\texttt{1.0.1}}
\newcommand{\releasecommit}{\texttt{feeccfb}}

\title{\agentname \hspace{2mm}1.0: Running Collider Simulations from Plain-Language Requests on Your Own Computer}

\author{
Aadarsh Singh \quad Sudhir K.~Vempati\\[0.5em]
\small Centre for High Energy Physics,
Indian Institute of Science, Bengaluru, India\\[0.3em]
\small
\texttt{aadarshsingh@iisc.ac.in}
\quad
\texttt{vempati@iisc.ac.in}
}
\date{August 2026}

\begin{document}
\maketitle

\begin{abstract}
We present \agentname{}, an open-source local interface for constructing a bounded class of
collider-simulation workflows from natural-language requests. A locally served language model
proposes a typed workflow representation, while deterministic software restores recognized
user-stated quantities, constructs the HEP-tool inputs, validates the supported workflow, and
presents the resulting artifacts for approval before execution. In a same-response comparison on
47 evaluable model--request cases, the first structured proposal produced 7 unmodified artifacts
satisfying the external benchmark scorer, compared with 19 after the complete deterministic
pipeline. After applying the benchmark's fixed representation normalization, the corresponding
counts were 11 and 43. The direction of improvement is unchanged; the gap between the two views
arises because the released builder and the benchmark scorers disagree on three bookkeeping
conventions (launch form, two fixed control lines, and the output-directory name), not on
physics content. Four of seven
approved workflows executed to completion on the managed local software stack with cross
sections consistent between repeats. In a separate challenge set, 57 of 96 problematic
requests nevertheless reached the approval stage after part of the request had been dropped,
defaulted, or reinterpreted. No tested unsafe payload was retained in an executable artifact
before the approval gate, but this does not establish operating-system-level containment. The
deterministic backend supports MadGraph, Pythia8, Delphes, and a restricted MadAnalysis~5 plan;
reliable natural-language routing to the MadAnalysis stage was not demonstrated in the tested
examples. Version~1.0.1 should therefore be viewed as an inspectable, validation-gated workflow
constructor requiring expert approval, rather than as an autonomous or scientifically
self-validating agent.
\end{abstract}

\section*{Program summary}
\begin{description}[leftmargin=1.9cm,style=nextline,itemsep=0.15em]
\item[\textit{Program title:}] \agentname
\item[\textit{Version:}] \agentversion{}
\item[\textit{Repository:}] \url{https://github.com/AadarshSingh0/HEPLocalAgent}
\item[\textit{Program files DOI:}] \url{https://doi.org/10.5281/zenodo.21934091}
\item[\textit{Licensing provisions:}] \agentname{} is distributed under the MIT License. MadGraph5\_aMC@NLO, ROOT, HepMC2, Pythia8, the MG5--Pythia8 interface, Delphes, MadAnalysis~5, and Ollama are separate components under their respective licenses.

\item[\textit{Programming language:}] Python 3.10 or newer, with a shell installer.

\item[\textit{External dependencies:}] MadGraph5\_aMC@NLO 3.5.13, ROOT 6.40.02, HepMC2 2.06.11, Pythia8 8.317, the MG5--Pythia8 interface 1.3, Delphes 3.5.1, and MadAnalysis~5 1.11.0. Ollama is used separately for model serving; \texttt{pydantic} is required by the Python package and \texttt{streamlit} only by the web interface.

\item[\textit{Nature of problem:}]
A collider-simulation workflow often passes through several programs,
each with its own syntax, configuration conventions, and output files.
Preparing this chain by hand can be time-consuming, and even a small
mistake may stop the run or alter the intended workflow. Language
models can make this interaction simpler, but their reliability varies
substantially across models and interfaces. Directly executing
model-generated commands also provides no independent check that the
requested process, physics model, collider energy, and simulation
stages have been preserved. This is especially relevant when using
models that can run locally on ordinary research hardware.
\item[\textit{Solution method:}]
\agentname{} separates language interpretation from workflow construction
and execution. The language model interprets the user's request and
produces a constrained description of the intended workflow, while
deterministic code checks explicit user requirements and the selected
physics model, constructs the required input files and commands for the
external programs, and presents the resulting workflow for approval
before execution. Requested outputs are checked after the run and the
workflow, validation results, and execution provenance are recorded.
For the full planning route, a limited number of model-guided revisions
can be attempted when validation fails.
\item[\textit{Restrictions:}]
\agentname{} is limited to the collider-simulation workflows represented
by its current schema and deterministic builders. Passing validation means
only that the constructed workflow satisfies the checks implemented by the
agent; it does not guarantee that every part of the original request has
been interpreted correctly, nor that MadGraph will find viable diagrams
or a nonzero cross section. The present release has no interactive
clarification turn, and the evaluation in Section~\ref{sec:evaluation} shows that some
problematic requests can reach the approval gate after parts of the
request are dropped, defaulted, or reinterpreted. Energy scans are limited
to 20 points and 200{,}000 requested events in total, and MadGraph
execution uses a fixed 1800\,s timeout. \agentname{} does not provide
operating-system-level sandboxing; external HEP programs run with the
filesystem permissions of the user who starts the agent.
The present release should not be used for unattended execution. Untrusted external inputs,
especially user-supplied UFO models, should be run in a restricted container or virtual machine.
\item[\textit{Running time:}] Planning time depends on model and hardware. On the benchmark host, Qwen3-Coder-Next had a warm median of about 0.40 minutes per benchmark request (P90 0.55 minutes), while Llama~3.3~70B had a median of about 3 minutes. Simulation time depends on the physical process, event count, enabled stages, and scan size; further timing details are given in Appendix~\ref{app:planner-selection}.
\end{description}

\section{Introduction}
\label{sec:need}

Collider studies commonly combine hard-process generation, parton
showering, detector simulation, and event analysis using tools such as
MadGraph5\_aMC@NLO, Pythia8, Delphes, and MadAnalysis~5
\cite{Alwall:2014hca,Sjostrand:2014zea,deFavereau:2013fsa,Conte:2012fm}.
A single study may therefore require the same physical setup to be
translated into several command formats, configuration cards, and file
conventions. Language models offer a natural way to simplify this
interaction. If both the model and the simulation software are run
locally, prompts, input files, and intermediate data can remain on
hardware controlled by the researcher. Local inference can also avoid
per-call charges and allow offline use once the required software has
been installed. The difficulty is that directly using model-generated
tool commands also allows an incorrect interpretation of the request to
become part of the simulation.

The companion HEPToolBench study~\cite{Singh:2026benchmark} examined
this problem at the level of individual HEP-software tasks. It found
that the choice of interface matters strongly. Models often struggled
when asked to produce native tool syntax directly, while the same
physical requests were handled much more reliably when the model
expressed its answer through a structured interface and ordinary code
constructed the final tool commands. \agentname{} builds on this
observation. The model is used to interpret what the user wants, while
deterministic software constructs and checks the workflow that can
actually be executed.
The exact procedure depends on the request. Simple process descriptions
can use a lightweight planning route, while richer requests use a typed
workflow representation. In both cases, explicit information supplied
by the user is preserved and the resulting workflow is checked against
the selected physics model and the features supported by the release.
The final tool inputs are then constructed by deterministic code and
shown to the user before execution. The model does not provide the shell
command, external-program launch sequence, or unrestricted analysis
code that is ultimately executed.

Several recent projects have explored language-model assistance and
agency in high-energy and astroparticle physics. FERMIACC and MadAgents
address theory construction and MadGraph-centred workflows
\cite{Agrawal:2026fermiacc,Plehn:2026madagents}. HEPTAPOD uses
schema-mediated coordination across multi-stage HEP tools, while GRACE
studies simulation-driven experimental design
\cite{Menzo:2025heptapod,Hill:2026grace}. DarkAgents employs a
multi-agent architecture for astroparticle-physics problems
\cite{Lucente:2026darkagents}. Some standardisation ideas are also being formulated \cite{Grosso:2026veraiphy}. These systems differ substantially in
scope, and local inference is not unique to \agentname{}; DarkAgents,
for example, explicitly documents Ollama support. The focus here is
different. \agentname{} deliberately restricts the role of the language
model and keeps workflow construction and execution within deterministic
software. This sacrifices some of the flexibility of more open-ended
agents, but makes the path from an interpreted request to the final HEP
tool inputs easier to inspect and test.

The practical question is then whether this separation actually helps
when the complete system is used with local models. We study this using
the released v1.0.1 software. We compare direct native-command
generation, structured serialization, and the complete \agentname{}
pipeline. We also test the execution of representative workflows and
examine deliberately challenging requests to identify cases in which
the implemented boundary is still incomplete.

\section{Design and implementation}
\label{sec:design}
The implementation follows three rules motivated by the companion
benchmark. First, the language model is used to interpret the request,
but its output is not treated as something that can be executed
directly. Second, deterministic code handles quantities and units,
constructs the required tool inputs, and launches the external
programs. Third, the resulting workflow must pass the implemented
validation checks before it can reach the execution gate.
This separation is central to the design of \agentname{}. The model is
allowed to interpret what the user means, while the part of the system
that constructs and executes the workflow remains ordinary, testable
software. Section~\ref{sec:evaluation} tests how well this design works
in the released version and where its checks remain incomplete.

\subsection{Architecture and runtime workflow}
\label{subsec:agent-runtime}

\begin{figure*}[!t]
    \centering
    \includegraphics[width=1.0\textwidth]{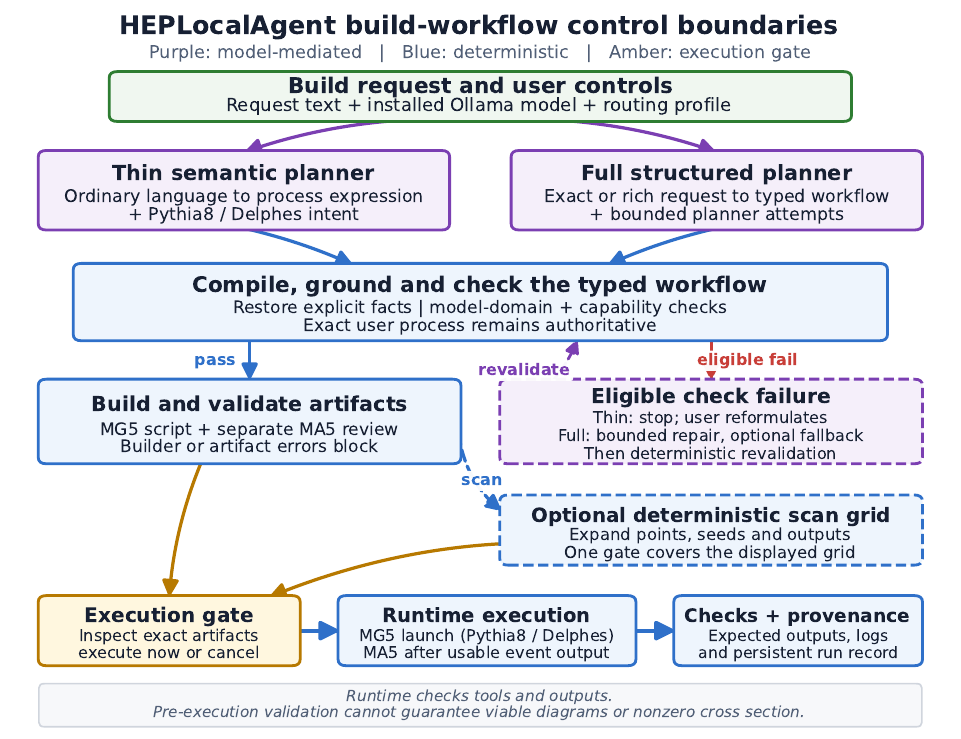}
    \caption{Architecture and control flow of \agentname{} in Build
    mode. Ordinary process descriptions follow the thin route, while
    explicit MadGraph processes and requests requiring structured
    analysis follow the full route. Only eligible full-route validation
    failures can enter bounded repair. After approval, execution is
    deterministic and the requested outputs are checked and recorded.}
    \label{fig:local-agent-workflow}
\end{figure*}

Figure~\ref{fig:local-agent-workflow} shows how a Build-mode request
moves from the user's text to an executed and recorded workflow. Chat
mode is separate and has no route to the HEP software. In Build mode,
the user chooses an installed Ollama model and a routing profile. The
model interprets the request. The profile sets operational choices such
as timeouts, repair limits, and an optional fallback model.

Build requests can follow two planning routes. Ordinary process
descriptions use the thin semantic route, which proposes a constrained
MadGraph process together with its interpretation of the requested
Pythia8 and Delphes stages. Requests containing explicit MadGraph syntax
or richer information, such as structured MadAnalysis~5 content, use
the full planner and its typed \texttt{WorkflowIntent}. Both routes
ultimately enter the same internal workflow representation. This keeps
the collider beams distinct from the incoming particles of the hard
process and distinguishes the total centre-of-mass energy from the
energy of each beam. It also records whether important values came from
the user, the model, or a default.

The planner output is not accepted as the final workflow. Information
stated explicitly by the user takes precedence over the model's
interpretation. The process, collider energy, event count, physics
model, and requested simulation stages are checked before the workflow
can proceed. On the full route, an explicit MadGraph process supplied
by the user is treated as authoritative rather than being silently
rewritten by the planner. The resulting workflow is then checked
against the selected physics model and the features supported by the
release.
Some full-route validation failures can be returned to the model for a
bounded repair attempt. The model receives the validation error and may
revise its proposal up to the limit set by the routing profile. Every
revision is checked again by deterministic code. Thin-route failures,
builder and artifact-validation failures, and runtime failures do not
enter this repair loop. The present release also has no interactive
clarification turn, so a thin-route request that cannot be handled must
be reformulated by the user. For installed or user-supplied UFO models,
particle and antiparticle names are extracted statically from
\texttt{particles.py} when possible, without importing or executing the
model. If a reliable particle namespace cannot be obtained, this check
remains permissive rather than rejecting a potentially valid
model-specific label.

The artifacts that reach execution are constructed by deterministic
builders. The complete MG5 command file is shown for inspection before
it is run, and a restricted MadAnalysis~5 plan is reviewed separately
when requested. Approval does not trigger another planning call. For an
energy scan, the individual scan points, seeds, and output names are
prepared before a single scan-level approval decision. Accepted
workflows are launched as subprocess argument lists without a shell and
with a fixed timeout.

The runtime checks look for the outputs requested by the workflow, not
only a successful process return code. Event generation must produce an
LHE file together with a parsed cross section and event count. A HepMC
file is additionally required when Pythia8 is requested, and a ROOT
file when Delphes is requested. Missing requested outputs are therefore
reported as failures even if an earlier part of the run completed.
MadAnalysis~5 is started separately when it was requested and usable
event output is available. Runtime failures are recorded but do not
trigger another model call.

Each workflow also produces a versioned JSON run record. It contains
the selected model and routing profile, the interpreted workflow,
validation and repair information, the approval decision, the exact
commands, and model-call timings. After execution, the record is
updated with quantities such as the return code, wall time, cross
section, event count, output paths, and requested-output checks. These
records make it possible to trace what the agent actually constructed
and executed. They do not establish that the requested physics is
meaningful, or guarantee that MadGraph will find viable diagrams or a
nonzero cross section.

\subsection{Model routing and supported workflow scope}
\label{subsec:agent-scope}

The package keeps the choice of language model separate from the routing
policy used around it. Table~\ref{tab:agent-profiles} lists the four
reference profiles included with the release. Each profile specifies a
primary model together with its timeout and repair settings, and may
also define a separate fallback model. In the web interface, the user
can replace the profile's primary model with another model available on
the configured Ollama host. The remaining routing settings are left
unchanged. If the selected primary and fallback are the same model, the
redundant fallback call is disabled.

Qwen3-Coder-Next is used as the reference primary because it gave a
useful balance between benchmark performance and response time on the
machine used in this work. This choice is not meant to define a
universal model ranking. The full quality--latency comparison is given
in Appendix~\ref{app:planner-selection}, together with the
agent-specific results obtained with the models used in
Section~\ref{sec:evaluation}.

\begin{table}[H]
    \centering
    \small
    \caption{Reference model-routing profiles included in the released
    source snapshot.}
    \label{tab:agent-profiles}
    \setlength{\tabcolsep}{5pt}
    \begin{tabularx}{\linewidth}{@{}l >{\raggedright\arraybackslash}X c >{\raggedright\arraybackslash}X@{}}
        \toprule
        Profile & Primary model & Max. primary repairs & Optional fallback \\
        \midrule
        \texttt{starter\_local} & \texttt{qwen2.5-coder:7b} & 2 & none \\
        \texttt{qwen\_primary} & \texttt{qwen3-coder-next:Q4\_K\_M} & 2 & none \\
        \texttt{qwen\_cascade} & \texttt{qwen3-coder-next:Q4\_K\_M} & 2 &
        \texttt{llama3.3:70b} (one repair) \\
        \texttt{legacy\_llama3} & \texttt{llama3:8b} & 3 & none \\
        \bottomrule
    \end{tabularx}
\end{table}

The implemented deterministic backend covers MG5 event generation, optional Pythia8
showering and hadronization, Delphes detector simulation, and
MadAnalysis~5 post-processing. Reliable natural-language routing is evaluated separately and is
not established for every backend capability. Delphes is used through the supported
MG5--Pythia8--Delphes chain; alternative shower generators are not
exposed in the present release. MadAnalysis~5 runs as a separate
post-processing step when a valid analysis plan is available. It uses
the most detailed event output produced by the workflow: ROOT after
Delphes, otherwise HepMC after Pythia8, or LHE for a parton-level run.
The MadAnalysis commands are constructed from the validated plan rather
than written freely by the language model.
Energy scans are expanded into separate runs, with a distinct seed and
output name for each energy point. The present release allows at most
20 scan points and 200{,}000 requested events across a scan. A
structured MadAnalysis~5 plan is limited to 20 histograms and 20 cuts.
These are operational guardrails intended to bound accidental
workloads; they are not physics limits. Representative requests and
their generated artifacts are shown in
Appendix~\ref{app:agent-examples}.

The released source is covered by a model-free suite of 440 unit,
regression, and acceptance-replay tests. These tests exercise the two
planning routes, validation and repair logic, deterministic artifact
construction, approval handling, output verification, and run
recording. The acceptance suite also replays representative model
outputs and checks whether they produce the expected workflow or are
blocked at the appropriate stage.
These software tests serve a different purpose from HEPToolBench. They
test how \agentname{} handles a proposed workflow once that proposal
has been produced. They do not measure how reliably a live language
model interprets a realistic request. That question is addressed in
Section~\ref{sec:evaluation}.

\subsection{Validation levels and security boundary}
\label{subsec:validation-levels}
\label{subsec:threat-model}
The term \emph{validation} is used at several stages of
\agentname{}, and these stages should not be confused with one another.
Table~\ref{tab:validation-levels} summarizes the distinction. A workflow
may, for example, be well formed and consistent with the implemented
checks without necessarily being a scientifically meaningful
calculation.

\begin{table}[H]
\centering
\small
\caption{Different levels of validation and verification considered in
\agentname{}.}
\label{tab:validation-levels}
\setlength{\tabcolsep}{4pt}
\begin{tabularx}{\linewidth}{@{}lY@{}}
\toprule
Level & Question addressed \\
\midrule
Schema validity &
Does the model output parse and satisfy the expected representation? \\
Intent grounding &
Are explicit choices such as the process, collider energy, event count,
physics model, and requested stages preserved? \\
Workflow validity &
Is the constructed workflow supported and internally consistent under
the implemented checks? \\
Runtime verification &
Did the external programs produce the outputs requested by the
workflow? \\
Scientific validation &
Does the resulting calculation correctly address the physics question
being studied? \\
\bottomrule
\end{tabularx}
\end{table}

\begin{figure}[H]
\centering
\includegraphics[width=0.60\linewidth]{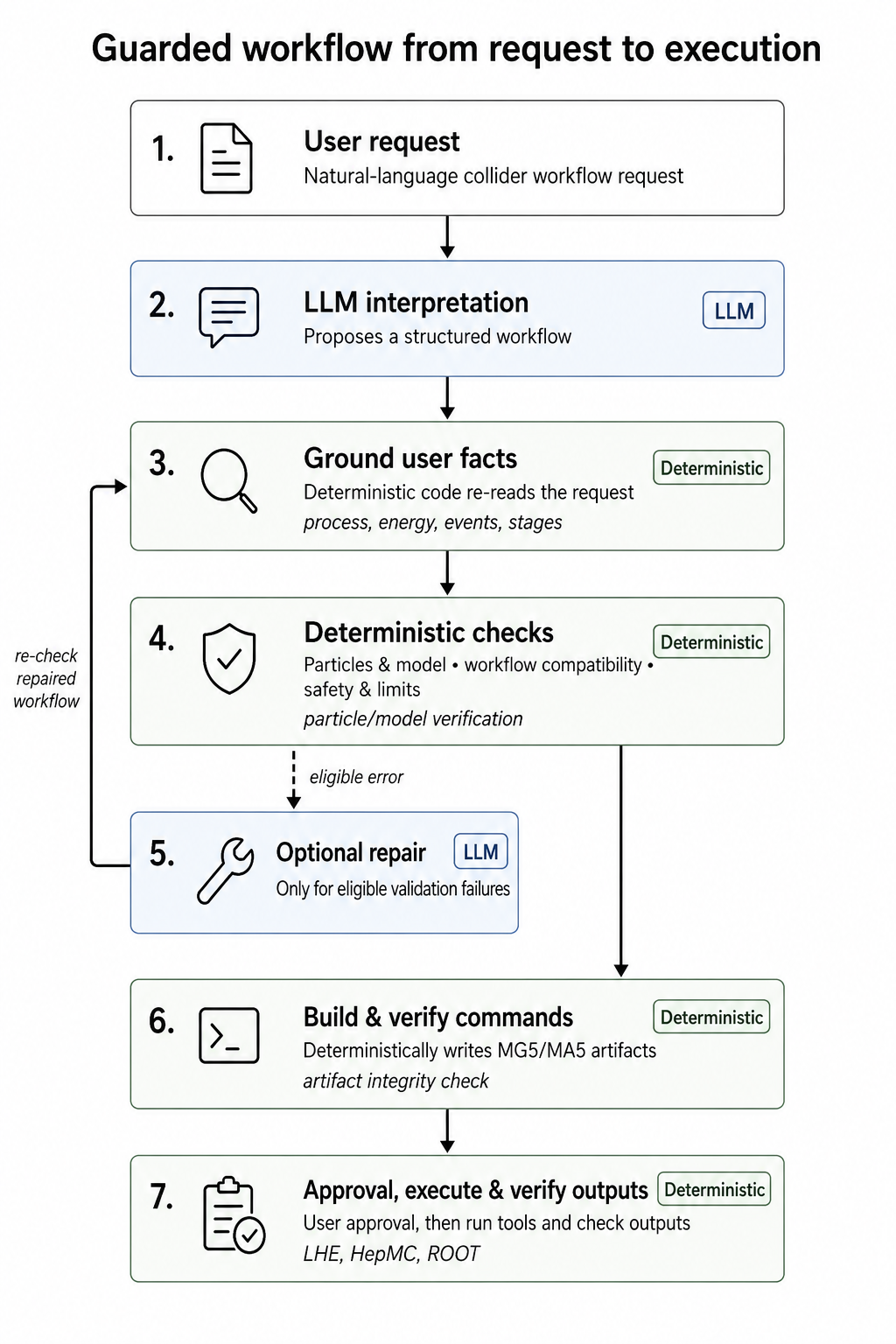}
\caption{Validation stages in a Build-mode request. Schema validity in
Table~\ref{tab:validation-levels} is checked on the model output at
step~2. Steps~3, 4, and 7 correspond respectively to intent grounding,
workflow validity, and runtime verification. Step~6 constructs the
artifact and checks its integrity, and step~7 additionally requires
user approval before execution. Grounding in step~3 restores explicitly
stated quantities that the deterministic parser can recognize.
Requirements expressed in freer language remain as the model
interpreted them. Only eligible full-route validation failures return
to the model for a bounded repair attempt, and every repaired workflow
is re-checked by the same deterministic path.}
\label{fig:validation-stages}
\end{figure}

\agentname{} provides software checks corresponding to the first four
levels. They occur at different points in the workflow and success at
one level does not imply success at the next; the fifth level lies
outside the software's scope.
Figure~\ref{fig:validation-stages} shows where these levels occur in a
Build-mode request. Table~\ref{tab:validation-levels} states what each
level asks, while Figure~\ref{fig:validation-stages} states where it is
enforced. Only two steps are model-mediated, the initial interpretation
and the optional bounded repair, and neither of them writes the artifact
that is finally executed.

The grounding step in Figure~\ref{fig:validation-stages} is not a
general re-interpretation of the request. It restores the quantities
that the deterministic parser can recognize with confidence: an explicit
MadGraph process expression, a stated collider energy with its units, an
event count, a named physics model, and explicitly requested simulation
stages. A user who writes the process in MadGraph syntax therefore has
that content carried through to the artifact unchanged. A user who
describes the same process conversationally is relying on the planner
model, and the grounding layer can then only confirm what the model
produced rather than recover the requirement independently. The reach of
the deterministic path is in this sense a property of how the request is
phrased. It is not a uniform guarantee, and the boundary cases studied
in Section~\ref{subsec:eval-containment} are concentrated in requests of
the second kind. This division of labour is motivated by the models the release targets.
The locally served models used here are small enough to run on ordinary
research hardware, and their performance on the companion benchmark
(Appendix~\ref{app:planner-selection}) varies substantially across
deployments. Asking such a model to check its own proposal would place
the reliability of the workflow back on the component whose reliability
is least assured, so the checks are kept in deterministic code that does
not depend on the model at all. A larger frontier model might well pass
many of these checks unaided, but it would not make the checks
inspectable, and it would not run on the researcher's own machine.

The same distinction is important when describing the execution gate.
Model-generated text is not executed directly. The inputs passed to the
HEP programs are constructed by deterministic code, and external
programs are launched without a shell. The user can inspect the
constructed workflow before execution. These choices provide a
controlled path from model interpretation to executable tool input, but
they should not be confused with operating-system sandboxing.
The external HEP programs run with the permissions of the user who
started \agentname{}. This also applies to user-supplied UFO models when
they are loaded by MadGraph. Untrusted code should therefore be run with
the usual additional isolation, for example in a container or virtual
machine under restricted permissions. Throughout this paper,
\emph{validation-gated} refers to the software checks and approval step
described above. It does not imply independent scientific validation or
system-level isolation. Section~\ref{subsec:eval-containment} tests how
well the implemented workflow checks preserve this boundary in
practice.

\section{Installation and usage}
\label{sec:install}

Each \agentname{} checkout keeps its Python environment and HEP
software stack within the repository directory. The Python environment
is stored in \texttt{.venv}, while ROOT, MadGraph5\_aMC@NLO, HepMC2,
Pythia8, the MG5--Pythia8 interface, Delphes, and MadAnalysis~5 are
installed under \texttt{.hep-stack}. Normal execution therefore does
not depend on HEP packages from the user's \texttt{PATH}, Conda or
Homebrew environments, another checkout, or a shared tools directory.
The complete installation used in this work was physically validated on
Ubuntu~24.04 x86-64 and Apple Silicon macOS.

The released version can be installed directly from the repository:
\begin{lstlisting}
git clone --branch v1.0.1 https://github.com/AadarshSingh0/HEPLocalAgent.git
cd HEPLocalAgent
./install.sh
\end{lstlisting}
The installation plan can first be inspected without modifying the
system using
\begin{lstlisting}
./install.sh --dry-run
\end{lstlisting}
The installer creates the managed HEP stack for the checkout and uses
pinned, checksum-verified downloads for its software components. The
installed versions and paths are recorded in the stack manifest.
Ollama is used separately for language-model serving and is not part of
the managed HEP stack. The agent can connect either to an Ollama server
on the same machine or to a configured remote endpoint. This allows the
HEP software and the language model to run on different machines when
needed.\\
After installation, the managed toolchain can be checked without making
a language-model call:
\begin{lstlisting}
.venv/bin/python -m hep_agent.doctor.cli --deep --timeout 180
.venv/bin/python -m hep_agent.selftest --events 1000
\end{lstlisting}
The self-test constructs a fixed $\proc{p p > e+ e-}$ workflow and runs
it through the managed MadGraph, Pythia8, Delphes, and MadAnalysis~5
chain. It checks the corresponding LHE, HepMC, ROOT, and analysis
outputs using the same managed paths used during normal execution.

The web interface is started with
\begin{lstlisting}
./run_agent.sh
\end{lstlisting}
In \emph{Build workflow} mode, the user describes the desired
calculation, selects a model and routing profile, and inspects the
interpreted workflow and generated tool inputs before deciding whether
to run it. After execution, the interface reports the run status and
available outputs.

The same workflow can also be run from the command line:
\begin{lstlisting}
.venv/bin/hep-local-agent run --profile starter_local \
  "simulate p p > t t~ in the Standard Model at 13 TeV with 10000 events, run Pythia8 and Delphes, save as ttbar_13tev"
\end{lstlisting}
The terminal interface displays the interpreted workflow and constructed
MG5 artifact before applying the same execution-approval policy.
Accepted runs use the managed HEP stack and produce a persistent run
record under \texttt{results/runs}.

\section{Evaluation of the released agent}
\label{sec:evaluation}
The model-free tests in Section~\ref{subsec:agent-scope} check how the
software behaves for controlled inputs and recorded planner outputs.
They do not measure how reliably a live model interprets a new
natural-language request, or how the complete system behaves near the
edge of its supported workflow space. We therefore evaluated the
released v1.0.1 software without adapting the agent to the evaluation
cases. Workflow-construction correctness was measured independently using the
pinned deterministic scorers from the companion HEPToolBench repository,
rather than the agent's own validation decisions.

\subsection{Protocol and baselines}
\label{subsec:eval-design}

The evaluation uses 20 valid requests from four workflow families:
Drell--Yan production, inclusive top-pair production, Higgs-plus-jet
production, and a top-pair workflow with Pythia8 and Delphes. Each
family is represented by five paraphrases that preserve the requested
physics while varying the wording. A further 30 requests probe the
boundary of the supported workflow space, giving 50 unique requests in
total. These include valid variants as well as invalid particle or
model choices, unsupported stage and scan combinations, ambiguous or
contradictory instructions, executable-code requests, excessive
resource requests, and validation edge cases. Their expected outcomes
were fixed before the live-model evaluation. The complete request set
and evaluation protocol are given in
Appendix~\ref{app:evaluation-protocol}.

We compare three ways of turning a request into a MadGraph workflow. In
\textbf{A (direct native)}, the model generates the native MadGraph
artifact directly. In \textbf{B (structured builder)}, the model
returns a typed workflow proposal and deterministic code constructs the
artifact, without the grounding and validation layer of the full
agent. In \textbf{C (full guarded agent)}, the proposal passes through
the complete \agentname{} grounding, validation, bounded-repair, and
execution-gate path. Conditions B and C use the same first usable model
response for each request. The response is replayed through both paths,
so their difference measures the effect of the deterministic agent
layer rather than variation between two model samples.

\begin{figure*}[t]
    \centering
    \includegraphics[width=\textwidth]{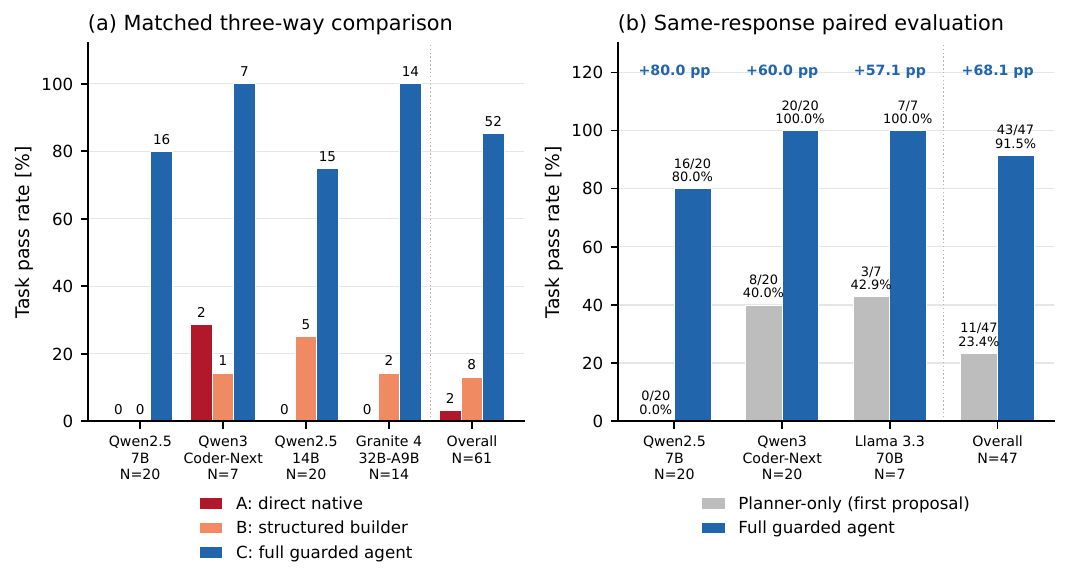}
    \caption{Workflow-construction evaluation of the released agent.
    (a) Matched comparison of direct native generation (A), structured
    intent with deterministic building (B), and the full guarded agent
    (C) on 61 units with semantic measurements in all three conditions.
    (b) Same-response comparison of the first structured proposal with
    the artifact produced after the full guarded pipeline on the 47
    semantic units of the primary study. Values shown use fixed representation normalization;
    the corresponding unmodified paired counts are 7/47 and 19/47. Task pass is determined by the
    corresponding external HEPToolBench scorer. Per-model denominators are shown beneath each
    group and differ because infrastructure-unavailable units are excluded rather than counted
    as semantic failures: in (a) the direct-native baseline limits Qwen3-Coder-Next to 7 units,
    and in (b) Llama~3.3~70B contributes 7 of 20 planned units.}
    \label{fig:agent-eval-conditions}
\end{figure*}

For the workflow-construction comparison, correctness is determined by
the corresponding deterministic scorer from the companion HEPToolBench
repository. The boundary study is evaluated separately against
predefined expected outcomes, distinguishing correct acceptance with
preserved intent from safe non-execution and false acceptance. The benchmark scorers expect a
small number of fixed artifact conventions. The evaluation therefore reports both unmodified
artifacts and a view after fixed representation normalization. Figure~\ref{fig:agent-eval-conditions}
uses the normalized view, while the corresponding unmodified counts are stated alongside it. The
normalization does not alter the process, collider energy, event count, physics model, requested
stages, or random seed. The exact rules are given in
Appendix~\ref{app:evaluation-protocol}.
A model-server or transport failure that occurs before any usable model
response is returned is classified separately as
\emph{infrastructure-unavailable}. Cases for which no usable model response was obtained are reported
separately and are not included in the workflow-correctness rates.

\subsection{Workflow-construction reliability}
\label{subsec:eval-paired}
\label{subsec:eval-baselines}

The paired B--C study planned 60 model--request cases and obtained
usable responses for 47 of them. The remaining 13 units are infrastructure-unavailable, all for Llama 3.3
70B, which therefore contributes 7 of its 20 planned cases. These cases
exhausted the three retry attempts at the 300\,s per-attempt transport
timeout. Llama 3.3 70B is the largest deployment tested here and is
close to the practical serving limit of the hardware used in this work;
its warm median response time on this host is about 3 minutes with a
P90 of 4.59 minutes (Table~\ref{tab:local-agent-quality-latency}), which leaves
little headroom against the timeout. The exclusion therefore reflects a
serving limit of the machine used for the evaluation rather than a
failure of the model to produce a workflow proposal, and a better-provisioned host would be expected to return more usable responses for
this model. On the 47 semantic units, the unmodified artifact from the
first structured proposal satisfies the external scorer in 7 cases (14.9\%), compared with 19
(40.4\%) after the complete pipeline. After fixed representation normalization, the corresponding
counts are 11 (23.4\%) and 43 (91.5\%). Under this normalized view, the paired transitions are 32
fail-to-pass, 11 pass-to-pass, 4 fail-to-fail, and no pass-to-fail.
The mean continuous score increases from 0.639 to 0.842. The direction of improvement is present
in both views. The gap between them has a specific cause: the released builder differs from the
benchmark contract in three bookkeeping conventions. It writes \texttt{launch <name>} where
the benchmark's workflow scorer requires a bare \texttt{launch}; it omits the \texttt{define p}
declaration and the \texttt{madspin=OFF} switch that the same scorer requires; and it names the
output directory from the request, whereas the top-pair and Higgs-plus-jet process scorers at
the pinned v1.2 commit require the fixed names \texttt{TTbar} and \texttt{HJ}, which the
paraphrased requests do not necessarily state. The companion benchmark's v1.2.1 correction
subsequently removed the \texttt{TTbar} requirement; the \texttt{HJ} name is stated in that
benchmark's own native prompt and is unchanged (Appendix~\ref{app:evaluation-protocol}). The three normalization operations (Appendix~\ref{app:evaluation-protocol})
are exactly the inverses of these differences. The unmodified count therefore measures agreement
with a convention the v1.0.1 builder was not written against, and the normalized count measures
satisfaction of the benchmark's checked physics and workflow fields after known representation
mismatches are removed; the artifact in Appendix~\ref{app:agent-examples} shows the builder's
form. Aligning the builder with the benchmark contract could make the two views coincide in a
future release.
This improvement does not come from model-guided repair. No unit in the
primary study entered the repair loop, whereas deterministic corrections
were applied in 36 of the 47 semantic units. As described in Section~\ref{subsec:validation-levels}, the
deterministic layer recovers only those requirements it can parse with
confidence, so condition~C is strongest exactly where the request is
explicit; Section~\ref{subsec:eval-containment} measures the
complementary case. These corrections are not restricted to syntax:
the grounding layer can restore an explicitly stated process, energy, event count, model, or
requested stage from the user request. The available aggregate records establish that 36 units
were corrected, but they do not include a field-level table distinguishing user-, model-, and
default-authority corrections. The B--C result should therefore be read as the joint effect of
deterministic grounding, correction, construction, and validation, not as an isolated measure of
better model reasoning; attribution of the gain to individual correction mechanisms remains
unresolved. The agent's execution gate is evaluated
separately. It disagrees with the external correctness
criterion in five cases. Four artifacts reach
\texttt{READY\_FOR\_APPROVAL} despite failing the external scorer,
corresponding to 4 of 46 ready units (8.7\%); all four share the same
placeholder user-UFO-path failure mechanism. In the opposite direction, one of the 43 scorer-passing full-agent
artifacts is blocked by an overly restrictive output-name check
(1/43, 2.3\%).

Detailed per-model results and both scoring views are given in Appendix~\ref{app:detailed-results}.
The matched three-way comparison addresses the broader effect of the
complete architecture. The 61-unit panel is the intersection of cases with usable outputs in all
three conditions; its composition is set mainly by the direct-native baseline, which obtained 63
usable responses from 80 planned, and comprises 20 units each for Qwen2.5-Coder 7B and
Qwen2.5-Coder 14B, 14 for Granite~4 32B-A9B, and 7 for Qwen3-Coder-Next. Under fixed
representation normalization, task-pass success is 2/61 (3.3\%) for direct native generation,
8/61 (13.1\%) for typed intent followed by deterministic building, and 52/61 (85.2\%) for the
full guarded agent. Seven units change from failure in A to success in B, one Qwen3-Coder-Next
unit passes in A but fails in B, and 44 change from failure in B to success in C. Overall, 50
units move from failure in A to success in C, and no unit that passes in A or B becomes a
failure in C.

This ordering is consistent with the interface effect observed in the
companion HEPToolBench study: introducing a structured representation
helps, but the much larger improvement is observed when the full
deterministic grounding and validation pipeline is added. The absolute pass rates should not be
compared directly across the two studies, and in particular the 8/61 obtained here for condition
B is not the same measurement as the 159/210 structured passes reported by HEPToolBench for its
matched pairs. The scoring target differs first: HEPToolBench scores the typed fields recovered
from the model's response against a prompt that names those fields, whereas condition B scores
the native artifact rendered by the release builder against the native scorer, including the
conventions discussed above. The requests also differ: HEPToolBench uses one fixed prompt per
task with a task-specific schema across 42 deployments, whereas condition B uses paraphrased
requests, the evaluation's general typed workflow proposal, and four locally served models. The
retained evaluation records do not attribute every proposal to a specific internal planner branch,
so the comparison does not identify a route-specific effect.
Condition B is therefore a harder and stricter test of structure alone than the benchmark's
structured condition; the two numbers are therefore not contradictory.
The comparison measures these layers as a whole and does not isolate the
contribution of each validator separately. The comparative runs also
used automatic approval, so the additional effect of human inspection
at the execution gate is not measured.

\subsection{Containment and failure analysis}
\label{subsec:eval-containment}
\label{subsec:eval-forensics}

Correct performance on valid requests does not show how the agent
behaves when the request itself is problematic. We therefore tested it
with additional requests containing invalid particle or model choices,
incompatible simulation stages, ambiguous or contradictory
instructions, excessive resource demands, and requests to execute
arbitrary code.
For valid supported requests, the agent preserved the requested physics
in 57 of 62 evaluable cases (91.9\%). None of the 62 valid-supported cases was rejected by the agent
(false rejection: 0/62); the remaining five failures instead reached approval without preserving the requested
physics. The agent's behaviour on problematic requests was substantially
weaker. Of 96 cases that should not have
proceeded to an executable workflow, only 39 were stopped before the
approval stage, while 57 reached approval in some modified form. We
count the latter conservatively as false acceptances. These requests
were deliberately chosen to probe difficult boundary cases, so the
59.4\% rate should not be interpreted as a failure rate for ordinary
use.

\begin{figure*}[t]
    \centering
    \includegraphics[width=\textwidth]{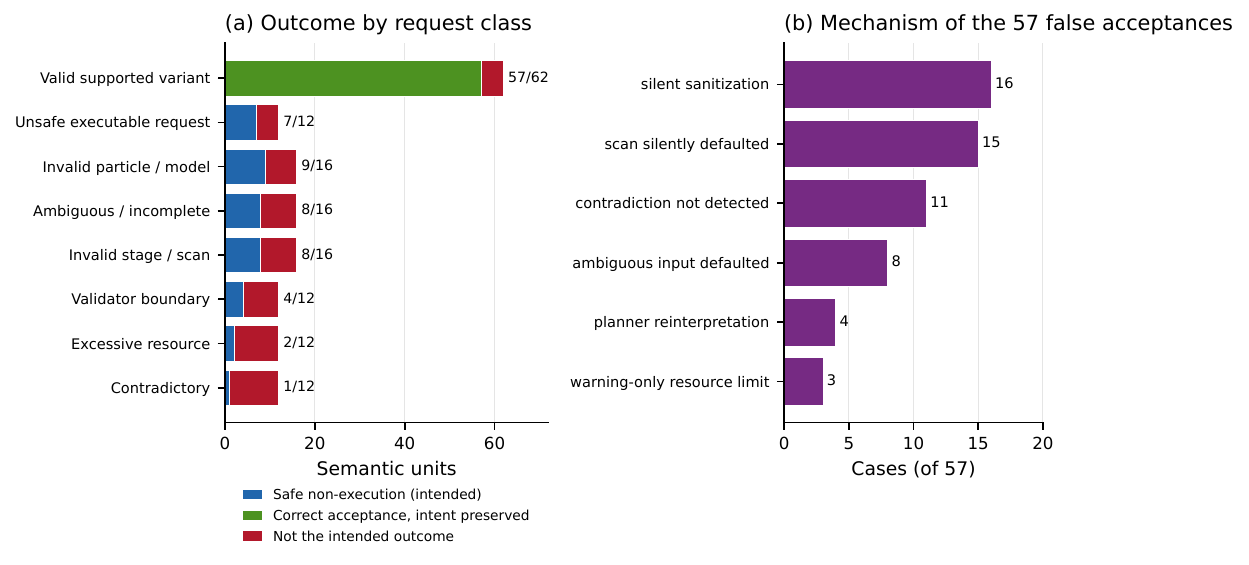}
    \caption{Behaviour of the released agent on valid and deliberately
    challenging requests. (a) For valid requests, the favourable outcome
    is acceptance with the requested physics preserved; for problematic
    requests, it is safe non-execution. The 62 valid supported units combine the six
    valid-variant boundary requests evaluated with four models and the 20 original requests
    evaluated with the two models added in the extension, less two units without a usable
    response. (b) Main mechanisms identified among the 57 of 96 problematic cases that reached
    approval and are conservatively counted as false acceptances.}
    \label{fig:agent-containment}
\end{figure*}

Importantly, no tested unsafe payload was retained in an executable artifact before the approval
gate. These cases were not approved and executed, and the result does not establish
operating-system-level containment. The
more common problem was that part of the user's request was silently
removed, defaulted, or reinterpreted before approval. In 17 cases this
changed an explicit part of the scientific intent. For scientific use,
this is the more important limitation: a workflow that fails visibly is
easier to identify than one that runs while answering a different
question.
The main failure modes are summarized in
Figure~\ref{fig:agent-containment}. They include accepting only a
sanitized part of the request, reducing a requested scan to a single
run, failing to detect contradictions, and choosing defaults for
ambiguous input. These results point to two clear priorities for future
versions: better contradiction handling and an explicit clarification
step when the user's intent is uncertain. Detailed class counts and
the full evaluation protocol are given in
Appendix~\ref{app:evaluation-protocol} and
Appendix~\ref{app:detailed-results}.

\subsection{Runtime and capability confirmation}
\label{subsec:eval-runtime}
\label{subsec:eval-capabilities}

Passing the workflow checks does not by itself guarantee that the
corresponding calculation will run successfully. We therefore executed
a representative subset of the accepted workflows with the managed HEP
software stack. Of seven tested workflows, four completed normally.
The repeated Drell--Yan calculations gave cross sections of 844.3 and
841.7\,pb, agreeing to 0.3\%, while two top-pair calculations gave
505.7 and 504.1\,pb, also agreeing to 0.3\%. The repeats used
independent random seeds, so this spread reflects Monte Carlo
statistical variation rather than a difference in the constructed
workflow.
The requested LHE, HepMC, and ROOT outputs were produced where the
corresponding simulation stages were enabled.
One larger top-pair sample also completed the physics simulation and
produced all requested outputs, but exceeded the fixed execution-time
limit during final processing and was therefore recorded as a timeout.
Two Higgs-plus-jet examples failed during MadGraph process generation
because $\proc{p p > h j}$ has no tree-level diagrams in the Standard
Model setup used here. The corresponding benchmark reference fails in
the same way. These examples illustrate an important distinction:
a workflow can be correctly constructed and pass the agent's checks
without the requested physics process being viable at runtime. The Higgs-plus-jet family is
therefore an artifact-construction test, not a successful end-to-end simulation.

We also performed a small six-case check of three capabilities that are
not covered well by the main Standard Model examples: energy scans,
user-supplied UFO models, and MadAnalysis~5. Both energy-scan requests
(2/2) and both UFO-model requests (2/2) were handled correctly. A
nine-point scan from 1 to 5\,TeV was successfully executed with an LHE
output at every point. As a UFO runtime example,
\texttt{MSSM\_SLHA2} was used to generate
$\proc{p p > x1+ x1-}$, giving
$\sigma = 0.4823 \pm 0.0026$\,pb for the tested setup.
Neither of the two MadAnalysis~5 requests preserved the requested
analysis stage (0/2). For the tested request to ``create default
plots'', both models produced an otherwise valid workflow but silently
omitted MadAnalysis~5. This is an intent-preservation failure rather
than a rejection: structured MadAnalysis~5 construction and its runtime
stage are implemented and covered by the model-free tests, but this
natural-language route to them was not interpreted correctly. MadAnalysis~5 is therefore a
restricted deterministic backend capability, not a reliably routed natural-language capability. The
six-case capability check is not included in the main performance
rates.

\subsection{Threats to validity}
\label{subsec:threats-validity}

The evaluation uses a fixed set of requests chosen to test specific
features of the agent, including several alternative phrasings of the
same underlying workflow. The reported success rates should therefore
be interpreted as results for these tests rather than as universal
performance estimates for all possible HEP requests. We also tested only a small set of locally served models, and the number of usable cases was not the same for every model.
The exact scores are sensitive to representation conventions: the paired full-pipeline
count changes from 19/47 for unmodified artifacts to 43/47 after normalization. As explained in
Sec.~\ref{subsec:eval-paired}, this reflects three bookkeeping conventions on which the v1.0.1
builder and the benchmark scorer disagree rather than errors in the physics content, but until
the builder is aligned with the scorer the unmodified count is the one an unmodified external
scorer would report. Human approval was
automated in the comparison experiments, so we do not measure how much
additional protection a physicist inspecting the workflow might provide. Model-guided repair was not exercised in the primary paired study and
occurred only once in the 158 evaluable cases of the extension, so its
independent contribution to end-to-end performance is not established
by this evaluation.
Finally, the runtime tests cover only a small number of representative
examples, while the problematic requests were deliberately chosen to
stress the limits of the system. These results should therefore be read
as evidence for the behaviour observed in our tests, rather than as a
prediction of how often problems will occur in ordinary use.

\section{Discussion and limitations}
\label{sec:outlook}

The evaluation supports the main design choice of \agentname{}:
separating language-model interpretation from deterministic workflow
construction substantially improves the reliability of supported
requests under both scoring views, although the absolute pass rate depends on whether the
artifact conventions emitted by the builder match those recognized by the external scorer
(Sec.~\ref{subsec:eval-paired}). The improvement is not primarily produced by repeated calls
to the language model. Instead, the deterministic layer uses information
from the user's request to construct and check the final tool inputs
before they can be executed.
The present release is deliberately limited in scope. It handles a
useful class of workflows involving MadGraph, Pythia8, and Delphes, together with a restricted
structured MadAnalysis~5 backend, but it does not attempt to represent every option
available in these programs. More complicated decay structures,
intermediate-state requirements, coupling constraints, and analysis
configurations remain only partly supported. The system is therefore
best viewed as general across processes that fit its present workflow
description, rather than as a complete natural-language interface to
all HEP software. Table~\ref{tab:benchmark-agent-crosswalk} summarizes how the failure modes observed
in the companion benchmark map onto the mechanisms implemented here and
the limitations that remain.

\begin{table}[t]
\centering
\small
\caption{Examples of HEPToolBench failure modes, the corresponding
mechanisms used in \agentname{}, and limitations that remain.}
\label{tab:benchmark-agent-crosswalk}
\begin{tabularx}{\linewidth}{@{}Y Y Y@{}}
\toprule
Benchmark failure & \agentname{} mechanism & Remaining limitation \\
\midrule
Invalid native syntax
& Deterministic artifact construction
& The structured intent itself can still be wrong \\

Incorrect beam or unit conversion
& Deterministic grounding and unit conversion
& The original request can still be misinterpreted \\

Unknown particle or model label
& Model-domain validation
& Checks can remain permissive when a UFO particle namespace cannot be established \\

Invalid simulation-stage combination
& Workflow validation
& Some valid but unsupported workflows remain outside the present schema \\

Malformed repair
& Bounded repair followed by deterministic revalidation
& Repeated semantic errors can still leave the request unresolved \\

Missing requested output
& Runtime output verification
& Successful output production does not establish scientific validity \\
\bottomrule
\end{tabularx}
\end{table}

The most important limitation observed in the evaluation is the handling
of ambiguous, contradictory, or only partly supported requests. In the deliberately challenging cases studied here, the current system
can drop or reinterpret part of the request instead of asking the user
for clarification. This is especially
important in scientific applications, where a technically valid
calculation may still be misleading if it answers a different question
from the one intended. An explicit clarification step, stronger
contradiction checks, and better preservation of scan and analysis
requirements are therefore priorities for the next release.
Until those mechanisms are implemented and evaluated, \agentname{} should not be used for
unattended execution. Expert approval is required, particularly for scans, ambiguous
specifications, contradictions, and requested analysis stages.

The checks performed before execution also have a clear boundary. They
can test whether a workflow is consistent with the request and with the
features represented by the agent, but they cannot guarantee that the
requested physical process has viable diagrams or that the resulting
cross section and distributions are scientifically meaningful. Those
questions remain the responsibility of the underlying HEP programs and,
ultimately, of the researcher. The current execution environment should
likewise not be regarded as an operating-system sandbox. User-supplied UFO models are code loaded
by MadGraph and should be handled with additional isolation when they are not trusted.

Future development will therefore focus first on making the
natural-language boundary more reliable rather than simply increasing
autonomy. Better clarification, richer but still structured workflow
descriptions, improved MadAnalysis routing, stronger resource limits,
and greater use of diagnostics from the HEP tools could broaden the
range of supported studies while keeping the final executable workflow
explicit and inspectable.

\section{Conclusion}
\label{sec:conclusion}

\agentname{} implements the central architectural lesson of HEPToolBench: a language model
interprets the user's request, while deterministic software restores recognized user-stated
facts, constructs the corresponding workflow, and checks it before execution. In the
47-case same-response comparison,
the unmodified structured artifacts pass the external scorer in 7 cases before the complete
pipeline and 19 afterward. Under the benchmark's fixed representation normalization, the
corresponding counts are 11 and 43. The direction of improvement is consistent across both
scoring views; the difference between them is a mismatch between the builder's launch form,
control lines, and output naming and the conventions the benchmark scorers require. This mismatch
could be removed in a future release by aligning the builder with the benchmark contract. The normalized
three-way comparison---2/61 for direct native generation, 8/61 for structured construction, and
52/61 for the complete pipeline---should be interpreted within the same evaluation contract.
The evaluation also identifies an important limitation. Difficult,
ambiguous, or contradictory requests are not always stopped. In the
challenge set, 57 of 96 requests that should not have proceeded still
reached the approval stage, usually after part of the original request
had been dropped or reinterpreted. No tested unsafe payload was retained in an executable
artifact before approval, but some cases changed the intended physics. A workflow can
therefore be technically valid while no longer representing exactly
what the user asked for.

The present release should consequently be viewed as an inspectable, validation-gated route
for constructing a bounded class of collider-simulation workflows requiring expert approval, not
as an autonomous replacement for scientific judgement or an operating-system sandbox. The most
important next step is to make the interaction with the user more
reliable when the request is incomplete or inconsistent, particularly
through explicit clarification and stronger preservation of scientific
intent. Within this scope, the results support combining language models with deterministic
grounding and workflow construction, while also showing that reliable preservation of scientific
intent remains the principal unresolved boundary.

\section*{Acknowledgements}

A.S. acknowledges the financial support and institutional resources provided by the Indian Institute of Science (IISc).
S.K.V. acknowledges the receipt of REDA funds from IISc.

\section*{Declaration of AI-assisted development and writing}

OpenAI's ChatGPT was used to assist with software implementation, debugging, documentation, manuscript drafting, and figure preparation. Anthropic's Claude was also used during
software development to review and improve parts of the agent and to propose code
patches. The authors defined the scientific objectives and software requirements,
reviewed and tested all AI-assisted code incorporated into the repository, executed
and inspected the reported workflows, and verified the technical and scientific
claims presented in the manuscript. The authors take full responsibility for the
software and the content of this work. Further details are recorded in the repository
file \path{AI_ASSISTED_DEVELOPMENT.md}.

\section*{Code availability}

The source code for \agentname{} is publicly available under the MIT License at
\url{https://github.com/AadarshSingh0/HEPLocalAgent}. The release described in
this work is v1.0.1 (commit \texttt{feeccfb}) and is permanently archived on
Zenodo~\cite{singh2026heplocalagent}. The external HEP programs and Ollama are
separate software packages distributed under their respective licenses.
The HEPToolBench benchmark and deterministic scorers used in the evaluation are
described in the companion paper~\cite{Singh:2026benchmark} and are available at
\url{https://github.com/AadarshSingh0/HEPToolBench}~\cite{HEPToolBenchRepo}; the reported
workflow-construction results use benchmark commit \texttt{509205c}, corresponding to
benchmark release v1.2, and have not been rescored under the corrected v1.2.1 scorers
reported there (Appendix~\ref{app:evaluation-protocol}). The manuscript source
contains the summary figures and tables needed to read the reported results.

\appendix

\section{Representative requests and artifacts}
\label{app:agent-examples}
The examples in this appendix illustrate how a natural-language request
is converted into the deterministic artifacts used by \agentname{}.
They are explanatory examples and are not additional evaluation cases.

\subsection{Single simulation with showering and detector simulation}

Consider the request:
\begin{quote}
``Simulate $pp\to t\bar t$ in the Standard Model at
$\sqrt{s}=13\,\mathrm{TeV}$ with 10{,}000 events. Run Pythia8 and
Delphes, and save the output as \texttt{ttbar\_13tev}.''
\end{quote}

The corresponding typed workflow represents the physics process,
collider energy, event count, requested simulation stages, and output
name as separate fields. The deterministic builder then converts these
fields into the native MadGraph artifact. In particular, the requested
$13\,\mathrm{TeV}$ proton--proton centre-of-mass energy is translated
to 6500\,GeV for each beam. The artifact presented for approval is

\begin{lstlisting}
import model sm
generate p p > t t~
output ttbar_13tev
set automatic_html_opening False --no_save
launch ttbar_13tev
shower=Pythia8
detector=Delphes
analysis=OFF
set nevents 10000
set iseed 0
set ebeam1 6500
set ebeam2 6500
done
\end{lstlisting}
Here the MadGraph process syntax and run commands are produced by the
builder rather than copied from free-form model output. Pythia8 and
Delphes are enabled because they were explicitly requested, while the
MadAnalysis stage remains disabled. Run settings not explicitly supplied
by the user retain the corresponding builder defaults. This artifact also shows the three
conventions on which the v1.0.1 builder and the companion benchmark's workflow scorer differ:
the builder writes \texttt{launch ttbar\_13tev} rather than a bare \texttt{launch}, and it
emits neither a \texttt{define p} line nor a \texttt{madspin=OFF} switch. MadGraph accepts the
artifact as written, but the benchmark scorer treats all three as pass-critical, which is why
Section~\ref{subsec:eval-paired} reports the paired results both with and without the
corresponding normalization.
After execution, \agentname{} checks for the outputs expected from the
requested stages and records them in the run record. For this workflow
these are the generated LHE event sample, the HepMC shower output, and
the Delphes ROOT file.

\subsection{Multiparticle final states and energy scans}

For an explicit MadGraph-style process, the particle content is
preserved directly by the deterministic builder. Examples include
\begin{lstlisting}
generate p p > e+ e- a
generate p p > j j a
generate p p > t t~ j
\end{lstlisting}
where \texttt{a} is the MadGraph label for a photon and the two
occurrences of \texttt{j} represent two final-state jet legs. This is
particularly useful when particle multiplicities or charge assignments
must be preserved exactly. The remaining uncertainty lies earlier in
the workflow: a conversational description such as ``two jets and a
photon'' must first be interpreted by the planner model. Once the
corresponding explicit process has been extracted and validated, the
builder preserves that particle content when constructing the final
MadGraph artifact.

Energy scans are handled separately. For a recognized request such as
``scan $pp\to\mu^+\mu^-$ from 1 to 5\,TeV in 500\,GeV steps'', the
start energy, endpoint, step size, and units are extracted
deterministically from the request. The nine-point grid
\[
1.0,\;1.5,\;2.0,\;2.5,\;3.0,\;3.5,\;4.0,\;4.5,\;5.0~\mathrm{TeV}
\]
is then constructed without asking the language model to enumerate the
individual energies. A validated base workflow is expanded across the
grid, with separate output names, random seeds, and run records for
each point.
The scan produces aggregate CSV and JSON summaries and, when usable
cross sections are available, a cross-section-versus-energy plot. The
released version accepts at most 20 scan points and 200{,}000 requested
events in total across the scan. It also rejects ranges for which the
endpoint cannot be reached by an integer number of steps, or for which
the required range information and units cannot be extracted.

These checks apply once a request has been recognized as an energy scan.
Natural-language recognition is a separate step and is not yet fully
reliable: in the boundary evaluation, some scan-like requests were
instead interpreted as ordinary single runs. The distinction is
important because a valid scan planner cannot protect a request that
never reaches the scan route.

\subsection{Examples blocked by implemented validators}

The following requests illustrate cases that the released validators
can block:
\begin{itemize}
\item ``Run Delphes without Pythia8'' --- rejected because the supported
detector-simulation path requires a preceding shower stage.
\item ``Simulate a 7\,TeV proton beam against a 4\,TeV proton beam'' ---
rejected because unequal beam energies are not represented by the
current workflow schema.
\item ``Execute the ROOT macro \texttt{analysis\_code.C} on the generated
events'' --- rejected because user-supplied executable analysis code is
not passed to external tools through the supported workflow.
\end{itemize}

These examples show specific checks implemented in the released
software. They should not be read as a guarantee that every alternative
wording of an unsupported request will be recognized and rejected. The
live boundary evaluation in Section~\ref{subsec:eval-containment}
measures that distinction directly.

\section{Evaluation protocol and scoring details}
\label{app:evaluation-protocol}
For each stage of the evaluation, the requests and their expected
outcomes were fixed before the corresponding live runs were performed.

\subsection{Request collection}

The primary set contains four workflow families linked to existing
HEPToolBench tasks, with five paraphrases of each request. The
paraphrases vary the wording while preserving the required physics.
Requests based on run-card lepton cuts are not included because the
workflow representation in v1.0.1 cannot express those cut fields.

\begin{table}[H]
\centering
\small
\caption{Primary request families and corresponding HEPToolBench tasks.}
\label{tab:eval-families}
\begin{tabularx}{\linewidth}{@{}Y c c Y@{}}
\toprule
Family & Paraphrases & Benchmark task & Main requirement \\
\midrule
Drell--Yan
& 5 & \texttt{mg\_basic\_001}
& $pp\to e^+e^-$ at the requested collider energy \\

Top pair
& 5 & \texttt{mg\_basic\_002}
& inclusive $pp\to t\bar t$ generation \\

Higgs + jet
& 5 & \texttt{mg\_basic\_003}
& $pp\to h j$ generation \\

Top-pair workflow
& 5 & \texttt{mg\_workflow\_005}
& $pp\to t\bar t$ with the requested run settings, Pythia8, and Delphes \\
\bottomrule
\end{tabularx}
\end{table}

A further 30 requests probe the boundary of the supported workflow
space. They are based on pre-existing unit tests, acceptance scenarios,
and documented boundaries in the released repository. Each request is
assigned a predefined expected outcome and source reference.

\begin{table}[H]
\centering
\small
\caption{Request classes in the 30-request boundary extension.}
\label{tab:eval-classes}
\begin{tabularx}{\linewidth}{@{}Y c Y@{}}
\toprule
Class & Requests & Expected outcome \\
\midrule
Valid supported variant
& 6 & Supported workflow with the stated intent preserved \\

Invalid particle or model
& 4 & Safe non-execution \\

Invalid stage or scan combination
& 4 & Safe non-execution \\

Ambiguous or incomplete
& 4 & Must not reach an executable ready state \\

Contradictory
& 3 & Safe non-execution \\

Unsafe executable request
& 3 & Safe non-execution; supplied Python, ROOT, or shell code must not execute \\

Excessive resource
& 3 & Safe non-execution beyond the documented resource boundary \\

Validator boundary
& 3 & Safe non-execution \\
\bottomrule
\end{tabularx}
\end{table}

Of the 30 boundary labels, 28 are unambiguous under the predefined
criteria and two are borderline cases. One concerns a
10-million-event request for which the released software issues a
warning rather than a hard rejection. The other contains a malformed
particle token that all four models interpret as the intended
$t\bar t$ final state. Both cases are retained under their original
expected outcomes for consistency.

\subsection{Evaluation conditions and model coverage}

Three evaluation conditions are used. In condition A, the model is
asked to produce the native MadGraph artifact directly. In condition B,
the model produces a structured workflow description that is converted
to the native artifact by deterministic code. Condition C uses the same
first structured model response as condition B, but passes it through
the complete \agentname{} grounding, validation, and construction
pipeline. Using the same response in B and C allows the effect of the
deterministic agent layer to be separated from variation between model
samples.\\
The primary paired study uses Qwen2.5-Coder 7B,
Qwen3-Coder-Next (\texttt{Q4\_K\_M}), and Llama~3.3~70B. The boundary
extension uses Qwen2.5-Coder 7B, Qwen3-Coder-Next,
Qwen2.5-Coder 14B, and Granite~4~32B-A9B. The extended evaluation contains 120 planned cases from the
30 additional requests evaluated with all four models, together with
40 cases from the 20 original valid requests evaluated with the two
additional models, Qwen2.5-Coder 14B and Granite~4~32B-A9B.
Of these 160 planned cases, 158 yielded usable model responses,
comprising 62 valid supported cases and 96 cases expected not to
proceed.

\begin{table}[H]
\centering
\small
\caption{Number of planned and evaluable model--request cases in each
part of the live evaluation.}
\label{tab:eval-coverage}
\begin{tabular}{@{}lrr@{}}
\toprule
Study & Planned & Usable model responses \\
\midrule
Primary paired comparison & 60 & 47 \\
Extended evaluation & 160 & 158 \\
Direct-native baseline & 80 & 63 \\
\bottomrule
\end{tabular}
\end{table}

The three-way comparison uses the 61 cases for which usable results are
available in all three conditions. A model--request case is included in
the corresponding performance calculation only when a usable first
model response is obtained. Transport or model-server errors are retried
up to three times with a 300\,s timeout per attempt. Failures in the
generated content itself, including physics errors, parsing failures,
or validator rejection, are treated as model outcomes and are not
retried.

\subsection{Scoring and normalization}

For the workflow-construction comparisons, correctness is defined by
the task-pass criterion of the corresponding HEPToolBench scorer at
benchmark commit \texttt{509205c}, which corresponds to benchmark
release v1.2. All workflow-construction counts reported in this paper
are therefore v1.2 quantities. The companion benchmark paper
subsequently corrected two native scorers in v1.2.1, so that
\texttt{mg\_basic\_002} accepts any syntactically valid output-directory
name and \texttt{mg\_workflow\_005} no longer charges the unstated
\texttt{analysis=OFF} and \texttt{done} conventions. Two of the four
request families evaluated here use those two tasks, and the first of
the two corrections overlaps the \texttt{rename\_output} normalization
defined below. Rescoring under v1.2.1 would therefore be expected to
raise the unnormalized counts and narrow the gap between the
unnormalized and normalized views; it would not affect the normalized
counts, which already apply that renaming. The results reported here
have not been rescored under v1.2.1. The boundary requests are evaluated
separately according to the expected outcomes defined in
Table~\ref{tab:eval-classes}.

The workflow-construction evaluation reports both unmodified artifacts and a view using fixed
representation normalization with three possible operations:
\texttt{rename\_output}, \texttt{bare\_launch}, and
\texttt{insert\_after}. Each operation undoes one difference between the v1.0.1 builder's
output and the benchmark contract. \texttt{rename\_output} replaces the output directory name
chosen from the request by the fixed name the corresponding benchmark scorer requires;
\texttt{bare\_launch} replaces the builder's \texttt{launch <name>} by the bare
\texttt{launch} the workflow scorer expects; and \texttt{insert\_after} adds the
\texttt{define p} declaration and the \texttt{madspin=OFF} switch, which the builder does not
emit and the workflow scorer treats as pass-critical. All three inserted or replaced forms are
accepted by MadGraph and do not change the requested process, particles, collider energy, event
count, physics model, simulation stages, or random seed. The same three operations applied to
the archived native responses of the companion benchmark change no task pass there; this is
consistent with their role as representation-level bookkeeping changes.

Random seeds are unchanged by this normalization. The unmodified and normalized
artifacts contain identical \texttt{iseed} values in all saved cases.
For the 11 evaluable top-pair-workflow cases, \texttt{set iseed 42} is
already present before normalization through the typed
\texttt{run.random\_seed} field. The deterministic grounding layer does
not independently reconstruct the seed from the natural-language
request, so correct seed handling still depends on the planner supplying
this field correctly. When no seed is supplied, the builder emits \texttt{set iseed 0} and
MadGraph draws a seed at run time; the run record then stores the
requested value rather than the seed actually used, so exact event-level
reproduction of such runs is not guaranteed by the record alone.
The unmodified artifacts quantify sensitivity to representation conventions. In the 47-case same-response study, the
first structured proposals pass in 7/47 cases before normalization and
11/47 after normalization. The corresponding full-agent artifacts pass
in 19/47 and 43/47 cases, respectively. The improvement produced by the
complete pipeline is therefore present in both views, although its magnitude depends strongly on
representation conventions. The main evaluation figure uses the normalized view.

The evaluation materials include the request sets, expected outcomes,
source-case mapping, model configurations, baseline prompts, routing
profiles, and cryptographic hashes identifying the versions used for
the reported results. The \agentname{} source corresponds to release
commit \releasecommit{}, and the HEPToolBench scorers use the benchmark
version specified above. Evaluation scripts and additional evaluation
tests are kept separate from the released production source. Runtime
checks replay the approved artifacts without further language-model
calls, with SHA-256 hashes used to verify that the executed files match
the approved files.

\section{Planner selection and latency}
\label{app:planner-selection}
\begin{figure}[H]
    \centering
    \includegraphics[width=0.98\linewidth]{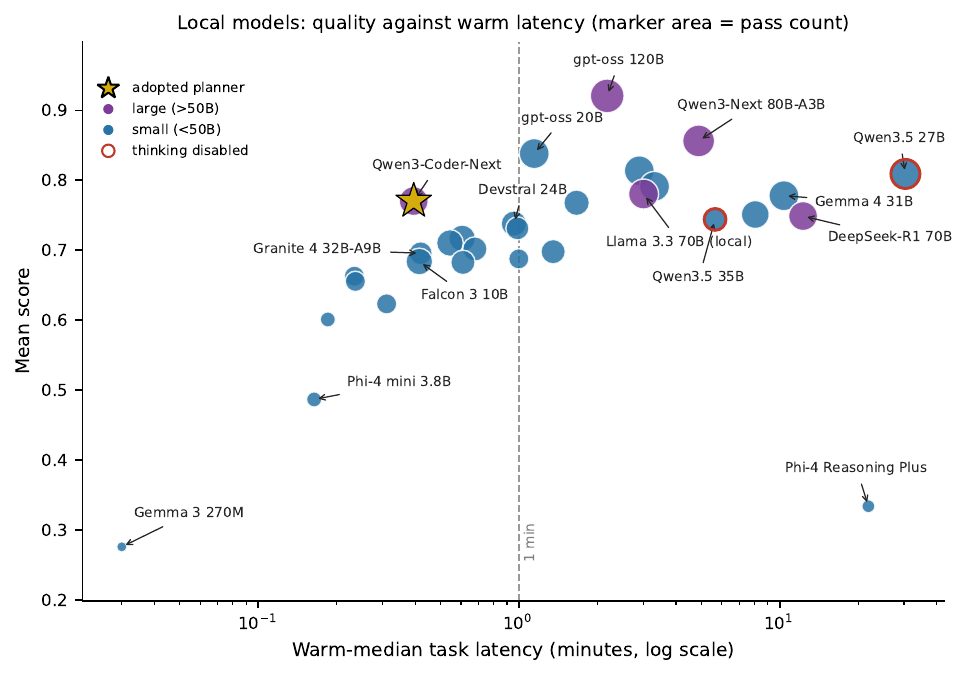}
    \caption{HEPToolBench quality versus warm per-task response time for
    the local-model cohort. Marker area indicates the number of passes
    on the 28-task main suite. The horizontal axis is logarithmic and
    the dashed line marks one minute. The starred point is the reference
    Qwen3-Coder-Next deployment; red rings denote Qwen3.5 configurations
    evaluated with thinking disabled.}
    \label{fig:local-quality-latency}
\end{figure}

The reference planner profiles were chosen by considering both
HEPToolBench performance and response time on the benchmark host.
Figure~\ref{fig:local-quality-latency} shows this comparison for the
local models. Quality is the single-run mean score on the 28-task main
suite. Latency is the wall-clock time measured through the Ollama
interface after excluding the first request, which reduces the effect
of model loading and other cold-start costs.
Qwen3-Coder-Next provided the most useful balance for this setup. The
tested \texttt{Q4\_K\_M} deployment obtained a mean score of 0.774 with
13/28 task passes and a warm median response time of 0.40 minutes.
Llama~3.3~70B scored slightly higher, at 0.782 with 14/28 passes, but
was considerably slower. These timings depend strongly on the hardware,
quantization, and serving configuration, so the choice of
Qwen3-Coder-Next should be understood as a practical reference choice
for the tested environment rather than a general ranking of models.

\begin{table}[H]
    \centering
    \small
    \caption{HEPToolBench performance and warm response time for the
    models used in the supplied planner profiles. Scores and passes refer
    to the 28-task main suite and are corrected v1.2.1 benchmark values, unlike the
    workflow-construction counts in the main text (Appendix~\ref{app:evaluation-protocol}).
    Times are in minutes and exclude the first timed request.}
    \label{tab:local-agent-quality-latency}
    \setlength{\tabcolsep}{4pt}
    \begin{tabularx}{\linewidth}{@{}Y c c c c@{}}
        \toprule
        Model & Passes & Score & Median & P90 \\
        \midrule
        \texttt{qwen3-coder-next:Q4\_K\_M} & 13 & 0.774 & 0.40 & 0.55 \\
        \texttt{llama3.3:70b} & 14 & 0.782 & 3.01 & 4.59 \\
        \texttt{qwen2.5-coder:7b} & 6 & 0.664 & 0.23 & 0.50 \\
        \texttt{llama3:8b} & 6 & 0.657 & 0.24 & 0.43 \\
        \bottomrule
    \end{tabularx}
\end{table}

For Gemma~4~31B, the plotted quality value is the canonical benchmark
result, while the latency is taken from a separate single-configuration
timing repeat because the canonical timing data pooled two context settings.

HEPToolBench performance does not by itself show how a model behaves
inside the complete agent. Table~\ref{tab:eval-planner} therefore
compares benchmark performance with the live \agentname{} results for
the four models used in the extended evaluation. The HEPToolBench
column reports the mean over five repeats of the full 31-task suite and
is therefore different from the single-run 28-task scores reported
above.
Among these four models, Qwen3-Coder-Next performs best both on the
original valid requests and on the boundary set. The sample is too small
to infer a general relationship between HEPToolBench score and agent
performance. In practice, planner selection should consider both
workflow correctness and response time, together with behaviour on
difficult or ambiguous requests.

\begin{table}[H]
    \centering
    \small
    \caption{HEPToolBench and live-agent performance for the four models used
in the extended evaluation. The HEPToolBench column is the mean score
over five repeats of the full 31-task suite. The valid-request column
reports correct workflow construction on the 20 original requests; the
Qwen3-Coder-Next and Qwen2.5-Coder 7B entries come from the primary
paired study of Section~\ref{subsec:eval-baselines}, while the
Qwen2.5-Coder 14B and Granite 4 entries come from the boundary
extension, for which Granite 4 has 18 evaluable cases. The boundary
column reports correct outcomes across the 30 boundary requests. The HEPToolBench
column reports corrected v1.2.1 benchmark values.}
    \label{tab:eval-planner}
    \setlength{\tabcolsep}{4pt}
    \begin{tabularx}{\linewidth}{@{}Y c c c@{}}
        \toprule
        Model & HEPToolBench mean & Valid requests & Boundary set \\
        \midrule
        \texttt{qwen3-coder-next:Q4\_K\_M} & 0.798 & 20/20 & 18/30 \\
        \texttt{qwen2.5-coder:14b}          & 0.747 & 15/20 & 13/30 \\
        \texttt{granite4:32b-a9b-h}         & 0.716 & 18/18 & 17/30 \\
        \texttt{qwen2.5-coder:7b}           & 0.673 & 16/20 & 15/30 \\
        \bottomrule
    \end{tabularx}
\end{table}

\section{Detailed evaluation results}
\label{app:detailed-results}
Table~\ref{tab:eval-paired} gives the model-level results of the
same-response comparison. The transition counts show how individual
cases change between the first structured proposal and the complete
\agentname{} pipeline.

\begin{table}[H]
    \centering
    \footnotesize
    \caption{Same-response comparison by model after fixed representation normalization. Transitions are ordered
    as fail$\rightarrow$fail, fail$\rightarrow$pass,
    pass$\rightarrow$pass, and pass$\rightarrow$fail. The final column
    gives the number of workflows that reached the approval stage but
    failed the external HEPToolBench scorer.}
    \label{tab:eval-paired}
    \setlength{\tabcolsep}{4pt}
    \begin{tabularx}{\linewidth}{@{}Y c c c c c c@{}}
        \toprule
        Model & $N$ & First proposal & Full agent & Change
        & Transitions & Ready but incorrect \\
        \midrule
        Qwen2.5-Coder 7B
        & 20 & 0/20 & 16/20 & $+80.0$ pp
        & 4/16/0/0 & 4/20 \\

        Qwen3-Coder-Next
        & 20 & 8/20 & 20/20 & $+60.0$ pp
        & 0/12/8/0 & 0/19 \\

        Llama 3.3 70B
        & 7 & 3/7 & 7/7 & $+57.1$ pp
        & 0/4/3/0 & 0/7 \\
        \midrule
        Overall
        & 47 & 11/47 & 43/47 & $+68.1$ pp
        & 4/32/11/0 & 4/46 \\
        \bottomrule
    \end{tabularx}
\end{table}
No case in this comparison entered the model-guided repair loop. For unmodified artifacts, the corresponding
pass counts are 7/47 for the first proposal and 19/47 for the full
agent. Thus the improvement remains present without normalization,
although its numerical size is smaller.
For the 96 evaluable cases expected not to proceed, safe non-execution is
observed in 1/12 contradictory cases, 2/12 excessive-resource cases,
4/12 validator-boundary cases, 8/16 ambiguous or incomplete cases,
8/16 invalid stage or scan cases, 9/16 invalid particle or model cases,
and 7/12 executable-code requests. The remaining 57 cases reach the
approval stage. Their main failure mechanisms are 16 cases in which
only a sanitized remainder of the request is retained, 15 scans reduced
to a single run, 11 undetected contradictions, 8 ambiguous requests
completed using defaults, 4 reinterpretations of malformed particle
tokens, and 3 requests exceeding a warning-level resource boundary.
These 57 cases stop at approval: none is approved for execution, and no tested unsafe payload is
retained in the executable artifact. This does not establish operating-system-level containment.
Nine are classified as low severity and 48 as moderate; none
is classified as high or critical severity.

The runtime checks use seven approved artifacts from the primary study.
Four complete normally. A 100{,}000-event top-pair workflow produces
all requested LHE, HepMC, and ROOT outputs but reaches the fixed
1800\,s executor limit during final bookkeeping. Two
Higgs-plus-jet workflows fail during MadGraph process generation because
$\proc{p p > h j}$ has no tree-level diagrams in the Standard Model
setup used here. Two additional runtime checks confirm the energy-scan
and \texttt{MSSM\_SLHA2} UFO examples described in
Section~\ref{subsec:eval-runtime}.

\begingroup
\small
\setlength{\bibsep}{2pt plus 0.3ex}
\bibliographystyle{unsrtnat}
\bibliography{refs}
\endgroup

\end{document}